\documentclass[12pt]{article}       
\usepackage{graphicx}

\begin{document}

\noindent{\Large{\textbf{Transition into a low temperature 
superconducting phase of unconventional pinning in Sr$_2$RuO$_4$}}} 
\vskip 1cm 
{\large {A.C. Mota$^{a,}$\footnote{Corresponding author.  
Present address: Laboratorium f\"ur Festk\"orperpkysik, ETH 
H\"onggerberg, CH 8093 Z\"urich, Switzerland.  Fax - +41 1 633 10 77 
E-mail: mota@solid.phys.ethz.ch} , E. Dumont$^{a}$, A. 
Amann$^{a,}$\footnote{Present address: UCSD, IPAPS-0360, La Jolla, 
California 92093-0360} , and Y. Maeno$^{b}$} \vskip 1cm 
\noindent{$^{a}$Laboratorium f\"ur Festk\"orperpkysik, ETH 
H\"onggerberg, 8093 Z\"urich, Switzerland}\\
$^{b}$ Department of Physics, Kyoto University, Kyoto 606-01, Japan }

\begin{abstract}

We have found a sharp transition in the vortex creep rates at a 
temperature 
$T^{\ast}=0.05 T_{c}$ in a single crystal of Sr$_2$RuO$_4$ ($T_{c}=1.03$\,K) by means of 
magnetic relaxation measurements. For $T < T^{\ast}$, the initial creep rates drop 
to undetectable low levels. One explanation for this transition into a 
phase with such extremely low vortex creep is that the low-temperature 
phase of Sr$_2$RuO$_4$ breaks time reversal symmetry. In that case, degenerate 
domain walls separating discreetly degenerate states of a superconductor 
can act as very strong pinning centers.\cite{sigrist}
\end{abstract}

Keywords: Ruthenates; Heavy Fermions; Superconductivity
\vskip 1cm 
The discovery of superconductivity in Sr$_2$RuO$_4$ in 
1994\cite{nature}, a material structurally similar to the high-$T_{c}$ 
superconductor (La$_{1-x}$Sr$_{x}$)$_{2}$CuO$_{4}$, provided the first 
example of a layered perovskite without copper which becomes 
superconducting.  The strong interest in this material is based on the 
suggestion\cite{rice} that Sr$_2$RuO$_4$ could constitute the first 
example of odd parity ($l=1$) superconductivity.  The suggestion was 
based on the fact that in the normal state above $T_{c}$, 
Sr$_2$RuO$_4$ behaves like a quasi-2D Landau Fermi liquid with 
many-body enhancements of the specific heat and the Pauli spin 
susceptibility similar to another Landau Fermi liquid, namely normal 
liquid $^3$He below about $T = 100$\,mK. At present, there is no 
experimental evidence for tripplet pairing in Sr$_2$RuO$_4$.  Some 
hints of unconventional pairing come from measurements of the specific 
heat\cite{maeno}.  In the cleanest samples it is found that in the 
superconducting state, the residual electronic specific heat remains 
at about 50\% of its normal value.
Furthermore, NQR measurements show no indication of a Hebel--Slichter 
peak in $1/T_{1}T$\cite{ishi} and also $T_{c}$ is strongly depressed by 
non-magnetic impurities\cite{mack}.  

Recently we investigated the magnetic properties of the unconventional 
superconductor UPt$_{3}$ by means of magnetic relaxation measurements 
on high quality single crystals\cite{andreas}.  We found out that in 
the low temperature B--phase, where a small spontaneous magnetic field 
has been observed in $\mu$SR experiments\cite{luke}, no creep could be 
detected from any metastable configuration for about the first 10$^4$ 
seconds.  Above the temperature at which the second jump in the 
specific heat occurs, we observed a different vortex regime.  In this 
regime, the initial vortex creep is finite with a rate that increases 
rapidly as the temperature approaches the transition temperature 
$T_{c}$.  We interpreted the zero initial creep rate in the 
low--temperature, low--field B--phase of UPt$_{3}$ as resulting from 
an intrinsic pinning mechanism where fractional vortices get strongly 
trapped in domain walls between domains of degenerate superconducting 
phases.

These experimental results show 
that the widely different pinning strengths can be used as an indirect 
information on the character of a given superconducting phase.  
Superconducting phases that break time reversal symmetry might then be 
identified by their lack of vortex creep or their anomalous strong 
pinning.  

\begin{figure}[h]
\begin{center}\leavevmode
\includegraphics[width=.8\linewidth]{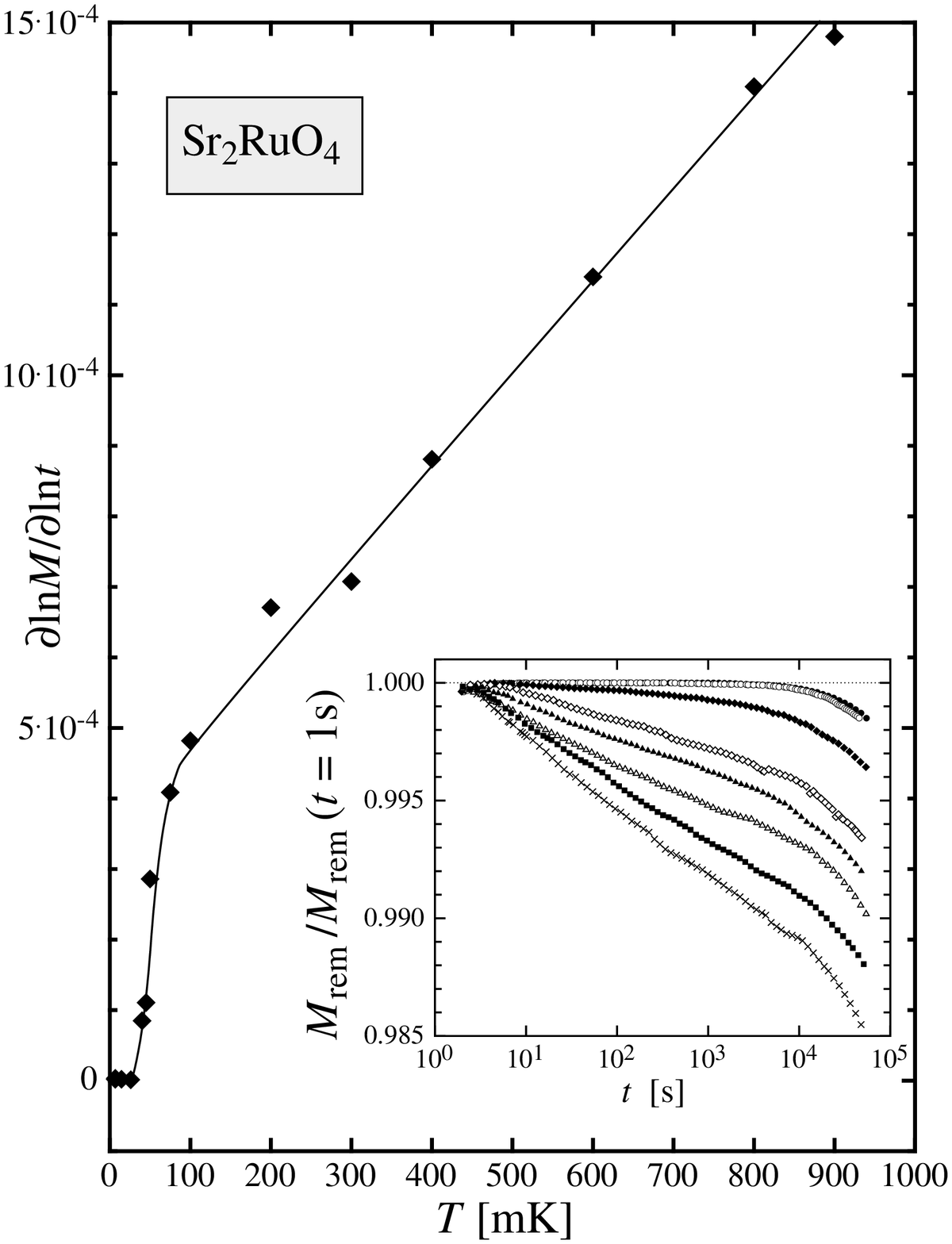}
\caption{Initial creep rates vs $T$. The inset shows decays of M$_{rem}$ 
for $T = 6.7$\,mK (closed circles), 26\,mK (open circles), 45\,mK  (closed 
diamonds), 100\,mK  (open diamonds), 200\,mK  (closed triangles ), 400\,mK  
(open triangles), 600\,mK  (closed squares), and 800\,mK  (crosses).  }
\label{decays}
\end{center}
\end{figure}

\begin{figure}[h]
\begin{center}\leavevmode
\includegraphics[width=.8\linewidth]{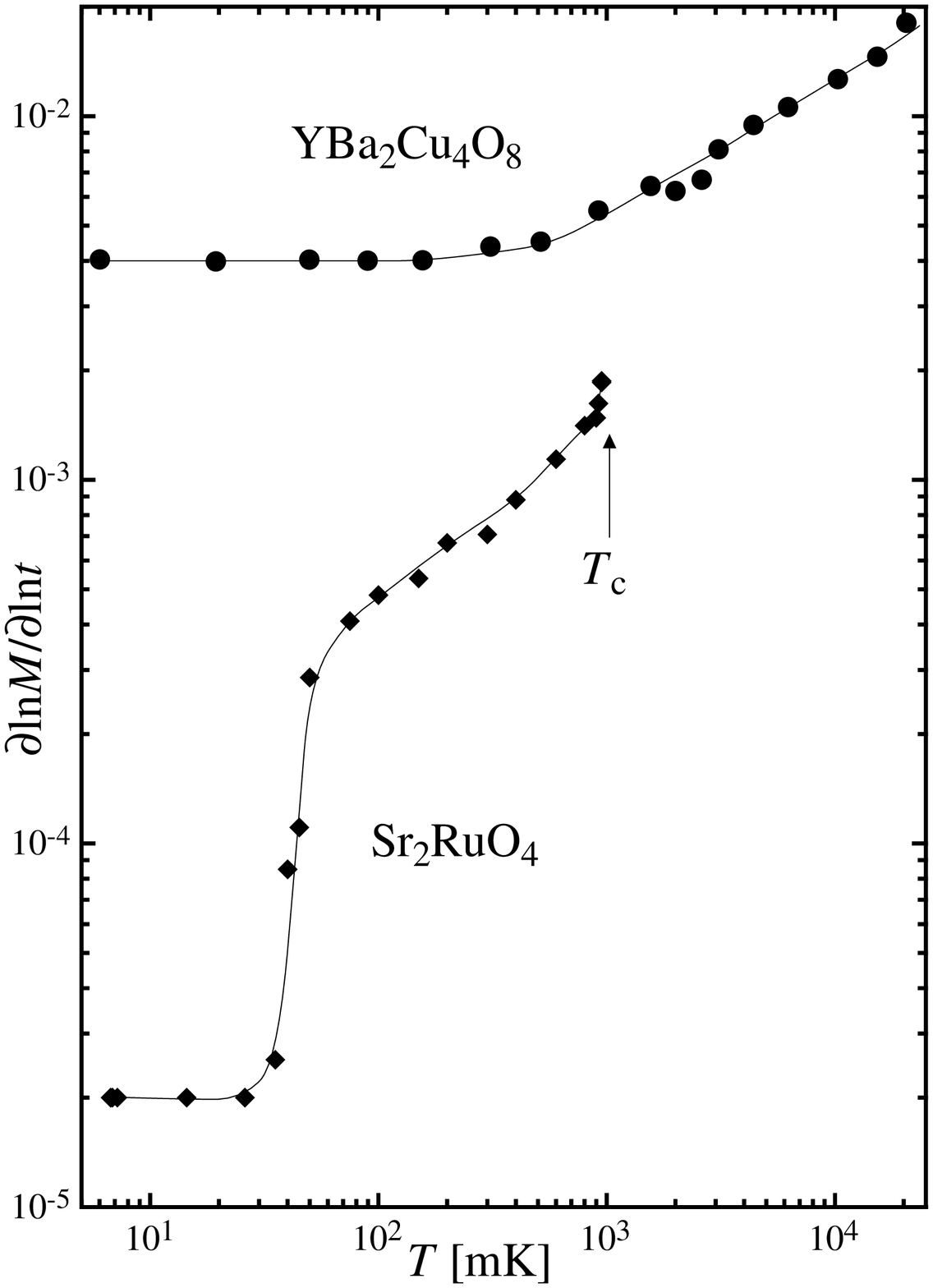}
\caption{Creep rates of YBa$_{2}$Cu$_{4}$O$_{7}$ and Sr$_2$RuO$_4$ vs $T$.}
\label{rates}
\end{center}
\end{figure}

We present here similar measurements of the relaxation of the remanent 
magnetization on a single crystal of Sr$_2$RuO$_4$.  The experimental 
arrangement has been described in a previous publication [7].  The 
single crystal has a transition temperature $T_{c}=1.03$\,K. The 
magnetic field in these measurements was applied at an angle of 
$15^\circ$ from the basal plane.  All the values of M$_{rem}$ were 
taken with the specimen cycled to sufficiently high fields, so that 
the sample was in the fully critical state.  In the insert of Fig.~1 
we give decays of the remanent magnetization normalized to the value 
of M$_{rem}$ at $t=1$\,s at different temperatures.  We observe that 
in the first couple of thousand seconds M$_{rem}$ relaxes following a 
logarithmic law.  At longer times one observes a more rapid 
relaxation, similar to what we found in UPt$_{3}$.  This long time 
rapid relaxation is due to surface vortices\cite{andreas}.  Here we 
only discuss the initial slopes of the decays which are determined by 
creep of bulk vortices.  The normalized creep rates $S_{initial}= 
\partial \ln M/\partial \ln t$ for Sr$_2$RuO$_4$ are given in Fig.~1 
as function of temperature.  We observe two different regimes of 
vortex creep separated by a rather sharp transition around $T\approx 
50$\,mK. For $T<50$\,mK the creep rates fall to zero within our 
sensitivity ($\partial \ln M/\partial \ln t \approx 10^{-5}$).  Above 
$T\approx 50$\,mK the creep rates are finite and increase rapidly as 
the temperature is increased.  In Fig.~2 we display the same data in a 
double logarithmic scale.  For comparison we have also plotted vortex 
creep rates of an YBa$_{2}$Cu$_{4}$O$_{8}$ single crystal\cite{mota} 
where the creep rates are much stronger and they tend to a finite 
value for $T\rightarrow 0$ on account of quantum tunneling.

Based on 
our previous result of no observable creep in the low temperature 
superconducting phase of UPt$_{3}$ we propose that the sharp 
transition in Sr$_2$RuO$_4$ at $T\approx 50$\,mK into a phase with 
very strong pinning might have a similar physical origin.  Work on 
more samples is under way to confirm these preliminary results.

\end{document}